\documentclass{article}
\pdfoutput=1
\usepackage[draft]{hyperref}
\usepackage[left=1.5in, right=1.5in]{geometry}
\usepackage{url}
\usepackage{microtype}
\usepackage{natbib}
\usepackage{amsmath}

\title{Is coding a relevant metaphor for building AI? \\\vspace{1em} \normalsize A commentary on ``Is coding a relevant metaphor for the brain?'', by Romain Brette}
\vspace{1em}
\author{Adam Santoro$^{*}$, Felix Hill\footnote{Equal Contribution}, David G. T. Barrett,\\ David Raposo, Matthew Botvinick, \& Timothy Lillicrap \\
\vspace{0.5em} \\
Deepmind, London UK\\
\small\texttt{\{adamsantoro,felixhill,barrettdavid,draposo,botvinick,countzero\}@google.com} \\
}
\date{}
\begin{document}
\maketitle
\begin{abstract}
Brette contends that the neural coding metaphor is an invalid basis for theories of what the brain does (Brette, 2019). Here, we argue that it is an insufficient guide for building an artificial intelligence (AI) that learns to accomplish short- and long-term goals in a complex, changing environment.
\end{abstract}
\vspace{1em}

The goal of neuroscience is to explain how the brain enables intelligent behaviour, while the goal of agent-based AI is to build agents that behave intelligently. Neuroscience, Brette attests, has suffered from an exaggerated (and technically inaccurate) concern for the codes transmitted by particular parts of the brain. In AI, on the other hand, some of the most notable recent progress has been made not by deeply considering neural coding and its implications, but by focusing on higher-level principles from optimization, learning, and control. 

Thanks to deep artificial networks trained via backpropagation, we now have artificial learning systems capable of impressive exhibitions of specific human-like skills, such as object recognition and language translation (e.g., He et al, 2016; Vaswani et al, 2017). In artificial, rather than biological, neural networks, we can more tractably characterize the relationship between a model’s neural codes, behaviour, and its external ‘world’. AI researchers have full access to their models' input data distribution, can visualise weights and activations in any part of the network and even make causal interventions on them, and can quickly implement new models informed by any coding hypotheses they may have. 

Nevertheless, in-depth analysis of a model’s internal representations is of increasingly rare concern for getting these models to work. Consider AlphaGo, which is one of the more compelling recent breakthroughs in AI (Silver et al, 2016). Researchers on this project precisely defined the model’s goals, the dynamics underlying the model’s interactions with its environment, how the model plans its actions, and how the model learns. Each of these components contributes to the model’s success, and yet none of them fundamentally depend on considerations from neural coding.

This is not to say that we cannot usefully apply representational analyses to such agents post-hoc, regardless of whether the representations satisfy Brette’s criteria for neural codes (Barack et al, 2019). Indeed, since the earliest days of connectionism researchers have been interested in the neural codes that emerge when a clearly-specified learning algorithm is applied to a well-understood model trained to execute a particular task. A more recent and important collaboration between AI and neuroscience revealed insight into the conditions under which well-known codes can emerge: grid-cells can increasingly be understood as the product of particular optimization processes (Banino et al, 2018; Cueva et al, 2018). A key feature of these examples, however, is the central descriptive role given to the learning algorithms, architectures, and optimization objectives; neural coding was incidental, and in many cases the codes were not fundamental, privileged primitives on top of which the models were built (Marblestone et al, 2016).

If the broader aim of agent-based AI (Russell et al, 2016) is to produce a system that accomplishes short- and long-term goals in a complex, changing environment, then there may be a more pernicious problem to the neural coding framework than it simply being out of vogue in modern AI. How internal responses arrive from given stimuli--a goal that is implicit in the neural coding metaphor--may be logically insufficient for producing intelligent behaviour. In outlining the reasons why, we recall arguments that any system--artificial or biological-- needs to exert control over its environment to achieve intelligent behaviour .

First, the observations with which an agent may compute do not exist as a prespecified dataset, independently of the agent's actions in the world. Rather, it is precisely the decisions that the agent takes in that world that determine the sensory data from which it learns. Second, “[w]ithout an ongoing participation and perception of the world there is no meaning for an agent” (Brooks, 1991). An agent participating in an external world that responds to its decisions learns useful, reliable, and meaningful interactions (Cisek, 1999). It is these meaningful interactions that ground the agent’s representations and allow them to be used for understanding and reasoning about its world. Therefore, insofar as neural coding is understood as a framework to help understand a system’s internal stimulus-response patterns, it is a logically insufficient framework for designing AI because of its failure to engage with the agent-environment causal loop.

Given these considerations, what, then, can we say about neural coding’s role in describing the brain? In neuroscience, we ultimately care about understanding how the brain enables intelligent behaviour. It is often argued that such an understanding cannot come from analyzing low-level, mechanistic details such as neural codes, because “[a] description of neural activity and connections is not synonymous with knowing what they are doing to cause behavior.” (Krakauer, 2017). For this level of understanding we need high-level computational and algorithmic theories that embrace agent-environment interactions. The history of AI tells us that the most useful principles, and the richest theoretical insights, emerged from studying control, optimization, and learning processes rather than the particularities of representations or codes (Sutton, 2019). A focus on inferring such processes using our increasing quantities of neural data, rather than characterizing neural codes for their own sake, may also be the most productive way of making progress on understanding intelligent behaviour in humans and animals.

\subsubsection*{Acknowledgments}
Thanks to Francis Song and Drew Jaegle for helpful comments. 

\subsection*{References}
Banino, A., Barry, C., Uria, B., Blundell, C., Lillicrap, T., Mirowski, P., ... \& Wayne, G. (2018). Vector-based navigation using grid-like representations in artificial agents. Nature, 557(7705), 429.

Barack, D. and Jaegle, A. (2019) Codes, Functions, and Causes: A Critique of Brette’s Conceptual Analysis of Coding

Brooks, R. A. (1991). Intelligence without reason.

Cisek, P. (1999). Beyond the computer metaphor: Behaviour as interaction. Journal of Consciousness Studies, 6(11-12), 125-142.

Cueva, C. J., \& Wei, X. X. (2018). Emergence of grid-like representations by training recurrent neural networks to perform spatial localization. arXiv preprint arXiv:1803.07770.

He, K., Zhang, X., Ren, S., \& Sun, J. (2016). Deep residual learning for image recognition. In Proceedings of the IEEE conference on computer vision and pattern recognition (pp. 770-778).

Krakauer, J. W., Ghazanfar, A. A., Gomez-Marin, A., MacIver, M. A., \& Poeppel, D. (2017). Neuroscience needs behavior: correcting a reductionist bias. Neuron, 93(3), 480-490.

Marblestone, A. H., Wayne, G., \& Kording, K. P. (2016). Toward an integration of deep learning and neuroscience. Frontiers in computational neuroscience, 10, 94.

Silver, D., Huang, A., Maddison, C. J., Guez, A., Sifre, L., Van Den Driessche, G., ... \& Dieleman, S. (2016). Mastering the game of Go with deep neural networks and tree search. nature, 529(7587), 484.

Russell, S. J., \& Norvig, P. (2016). Artificial intelligence: a modern approach. Malaysia; Pearson Education Limited.

Sutton, R. (2019). The Bitter Lesson.  Available at \url{http://incompleteideas.net/IncIdeas/BitterLesson.html}

Vaswani, A., Shazeer, N., Parmar, N., Uszkoreit, J., Jones, L., Gomez, A. N., ... \& Polosukhin, I. (2017). Attention is all you need. In Advances in neural information processing systems (pp. 5998-6008).

\end{document}